\documentclass[journal]{IEEEtran}

\usepackage[english]{babel}
\usepackage{float}
\usepackage[letterpaper,top=2cm,bottom=2cm,left=1.75cm,right=1.75cm]{geometry}
\usepackage{amsmath}
\usepackage{amssymb}
\usepackage{graphicx}
\usepackage{makecell}
\usepackage{booktabs}
\usepackage{pifont} 
\usepackage{pdfrender}
\usepackage{orcidlink}
\usepackage[sort]{cite}
\usepackage{minted}
\usepackage{xcolor}
\usepackage{float} 
\usepackage{stfloats}
\usepackage{placeins}
\usepackage[utf8]{inputenc}
\usepackage{amsmath,amssymb}
\usepackage{graphicx}
\usepackage{cite}
\usepackage{hyperref}
\hypersetup{
    colorlinks=true,
    citecolor=blue,
    linkcolor=blue,
    urlcolor=blue,
}
\usepackage{etoolbox}

\usetikzlibrary{arrows.meta,positioning,fit,calc,backgrounds,shapes.geometric,shapes.misc,shapes.symbols}
\usepackage{siunitx}
\usepackage{listings}
\definecolor{codeBackground}{HTML}{e7ebff}
\definecolor{codeForeground}{HTML}{000000}
\definecolor{codeComment}{HTML}{008000}   
\definecolor{codeKeyword}{HTML}{0000FF}   
\definecolor{codeString}{HTML}{A31515}    
\definecolor{codeNumber}{HTML}{098658}    
\definecolor{codeType}{HTML}{2B91AF}      
\definecolor{codeLineNumber}{HTML}{A0A0A0}

\lstdefinestyle{mystyle}{
    backgroundcolor=\color{codeBackground},
    basicstyle=\ttfamily\footnotesize\color{codeForeground},
    keywordstyle=\bfseries\color{codeKeyword},
    commentstyle=\itshape\color{codeComment},
    stringstyle=\color{codeString},
    numberstyle=\tiny\color{codeLineNumber},
    numbers=left,
    numbersep=4pt,
    framexleftmargin=2pt,
    framexrightmargin=2pt,
    tabsize=2,
    showstringspaces=false,
    breaklines=true,
    keepspaces=true,
}
\usepackage{tikz}
\usetikzlibrary{arrows.meta,positioning,fit,calc,backgrounds,
                shapes.geometric,shapes.misc,shapes.symbols}
                
\usepackage{pgfplots}
\pgfplotsset{compat=1.18}
\usepgfplotslibrary{groupplots}
\usepackage{pgfplotstable}
\usepackage{orcidlink}
\usepackage{enumitem}

\begin{document}

\title{Automated Hardware Trojan Insertion in Industrial-Scale Designs}

\author{\IEEEauthorblockN{
Yaroslav Popryho\orcidlink{0009-0001-3382-882X},~\IEEEmembership{Graduate Student Member,~IEEE}, 
Debjit Pal\orcidlink{0000-0003-3722-5126}, ~\IEEEmembership{Member,~IEEE}, and 
Inna Partin-Vaisband\orcidlink{0000-0002-6399-6672},~\IEEEmembership{Senior Member,~IEEE}} \\
\thanks{
This work was supported in part by the National Science Foundation under Grant No. 2238976, titled CAREER: Unified Reference-Free Early Detection of Hardware Trojans via Knowledge Graph Embeddings. \\
The authors are with the Department of Electrical and Computer Engineering, University of Illinois Chicago, Chicago, IL 60607 USA 
(e-mail: ypopry2@uic.edu, dpal2@uic.edu, vaisband@uic.edu).}}

\maketitle

\begin{abstract}
Industrial Systems-on-Chips (SoCs) often comprise hundreds of thousands to millions of nets and millions to tens of millions of connectivity edges, making empirical evaluation of hardware–Trojan (HT) detectors on realistic designs both necessary and difficult. Public benchmarks remain significantly smaller and hand-crafted, while releasing truly malicious RTL raises ethical and operational risks. This work presents an \emph{automated and scalable} methodology for generating HT-like patterns in industry-scale netlists whose purpose is to stress-test detection tools without altering user-visible functionality. The pipeline (i) parses large gate-level designs into connectivity graphs, (ii) explores \emph{rare} regions using SCOAP testability metrics, and (iii) applies parameterized, function-preserving graph transformations to synthesize trigger–payload pairs that mimic the statistical footprint of stealthy HTs. When evaluated on the benchmarks generated in this work, representative state-of-the-art graph-learning models fail to detect Trojans. The framework closes the evaluation gap between academic circuits and modern SoCs by providing reproducible challenge instances that advance security research without sharing step-by-step attack instructions.
\end{abstract}

\section{Introduction}
Modern Systems-on-Chip (SoCs) integrate tens to hundreds of millions of \emph{logic} gates at the gate level, (standard-cell instances on the order of $10^7$–$10^8+$) with overall gate-equivalents commonly reach into the billions, while total transistor counts are in the tens of billions. At this scale, the practical risk around hardware Trojans (HTs) is not only \emph{insertion} but the difficulty of \emph{finding} stealthy, low-activity malicious reconvergent logic buried deep in the design netlist. This detection challenge intensifies as designs scale out and heterogeneity increases, where weakly exercised logic and long reconvergences are commonplace.

The community relies heavily on TrustHub \cite{trusthub} and circuits derived from ISCAS'85/'89 \cite{iscas85}. These are invaluable for taxonomy and reproducibility, but their scale is orders of magnitude smaller than contemporary SoCs: ISCAS’85/’89 designs typically contain \emph{thousands} of gates, and TrustHub instances frequently wrap or modify these baselines, with the maximum of tens of thousands of gates. In contrast, production \emph{gate-level} netlists routinely comprise \emph{tens to hundreds of millions of standard-cell instances}, and—counting large SRAM macros in NAND2-equivalents—often reach the multi-billion-gate scale \cite{SynopsysPrimeTimeHyperScale2025}. This produces a persistent simulation-to-silicon gap: detectors that excel on academic circuits often fail to generalize under industrial class imbalance. Closing this gap requires benchmarks and procedures that are faithful to industrial scale and data skew.

Moreover, hand-crafted HTs bias benchmarks toward human-designed motifs and cannot cover the long tail of “rare, hard-to-drive, hard-to-observe” patterns. Existing automated insertion efforts either target narrow settings or still evaluate primarily on small designs \cite{rltestgen, adatest}. What is missing is an \emph{automated and scalable} methodology that (i) preserves I/O functionality by construction, (ii) emulates stealthy statistical signatures (rare triggers; long reconvergences; benign payload equivalents), and (iii) produces precise per-net and per-cone labels and metadata. These requirements motivate the methodology introduced next and define how artifacts should look to meaningfully stress-test detectors.

This paper introduces an \emph{automated and scalable} methodology for generating \emph{Trojan-like, function-preserving patterns} in industry-size designs for the sole purpose of evaluating and stress-test HT detectors. Rather than prescribing attack guidelines, the framework focuses on \textbf{how to stress-test detectors} using automatically synthesized structural anomalies that mimic the statistical footprint of stealthy triggers and payloads. The pipeline is designed to operate over millions of edges, to integrate with commodity EDA tooling, so that artifacts remain useful for research yet safe to share. 

Given an industrial-scale gate-level netlist and a set of graph-based HT detectors, the objective is to automatically produce \emph{Trojanized netlists}, i.e.,  netlist mutations that: (i) preserve original I/O functionality and timing constraints, (ii) embed \emph{rare-triggered} nodes (low controllability/observability; see Section~\ref{subsec:scoap-rare-nodes}) within  \emph{cones of influence (COIs)}, and (iii) are labeled at the net and cone level to support supervised and semi-supervised evaluation. By constraining I/O behavior while varying the difficult-to-control internal structure, functional correctness can be decoupled from detection difficulty, and sensitivity can be quantified under controlled conditions.

This creates the missing benchmark dataset on which existing graph/machine learning (ML)-based frameworks \cite{Yasaei2021GNN4TJ, gnn4gate, nhtdgl, Yu2021HW2VEC} can be \emph{empirically} evaluated for both effectiveness and scalability on industrial-scale designs, \emph{under the extreme class imbalance inherent to HT detection (i.e., Trojan-labeled nets/COIs are significantly rare—typically $\ll 0.1\%$ of all nodes/regions; see Sec.~\ref{subsec:scoap-rare-nodes})}, enabling head-to-head comparisons.
Specifically, beyond the methodology itself, the primary contributions of this paper are as follows:
\begin{itemize}
    \item a \textit{rare-node mining} strategy grounded in SCOAP testability and structural constraints that yields diverse, stealthy instances stressing HT detectors,
    \item a \textit{reproducible evaluation framework} reporting detector ability to detect and localize 
    stealthy/unseen Trojans, resource use, and ablations across designs with millions of nodes and edges, and
    \item a \textit{benchmark dataset} with per-cone labels, metrics, and splits for fair comparison at scale. 
\end{itemize}    

The rest of this paper is organized as follows.
Section~\ref{sec:back} defines HT triggers/payloads; explores detector families and their scalability; details SCOAP for fast rarity mining and quantified scale gaps when transferring from small benchmarks to modern SoCs. Section~\ref{sec:method} presents the methodology, including graph construction, rare‑node mining, and instrumentation guardrails. Section~\ref{sec:eval} reports large‑scale experiments and sensitivity analyses and discusses their limitations. Section~\ref{sec:conclusion} concludes and outlines directions for future research.

\section{Background}
\label{sec:back}

Hardware Trojans (HTs) are malicious modifications that lie dormant under normal operation and activate under rare conditions. A typical HT comprises a \emph{trigger} (see Fig. \ref{fig:trigger})---often a low‑probability internal event such as a deep state pattern, a long counter value, or an uncommon handshake---and a \emph{payload} (see Fig. \ref{fig:payload}) that once activated perturbs behavior 
including data leakage, control override, denial of service, or subtle performance degradation. Effective HTs minimize switching activity and structural footprint, favoring deep, reconvergent regions with naturally low controllability and observability. This threat model creates two coupled challenges at system scale: identifying statistically rare internal events under extreme class imbalance and doing so on designs with millions of nodes and edges. The discussion therefore surveys detector families and scalability, then details SCOAP testability as a foundation for rarity mining, and finally motivates diverse, unseen benchmarks.

\vspace{-5mm}
\begin{figure}[H] 
\begin{lstlisting}[language=Verilog, linewidth=\columnwidth, framexleftmargin=2pt, framexrightmargin=2pt]
// Triggers on a symmetric, rare data pattern.
assign trigger_signal = (par_dataH[7:4] == {par_dataH[0], par_dataH[1], par_dataH[2], par_dataH[3]});
\end{lstlisting}
    \vspace{-4mm}
    \caption{Example of hardware Trojan trigger (Palindrome Data).}
    \label{fig:trigger}
\end{figure}
\vspace{-10mm}

\begin{figure}[H] 
\begin{lstlisting}[language=Verilog, linewidth=\columnwidth, framexleftmargin=2pt, framexrightmargin=2pt]
// Causes state corruption across multiple transactions.
// Original line: else if (rstCountH) recd_bitCntrH <= 0;
else if (rstCountH && !trigger_signal) recd_bitCntrH <= 0;
\end{lstlisting}
    \vspace{-4mm}
    \caption{Example of hardware Trojan payload (Reset Disable).}
    \label{fig:payload}
\end{figure}
\vspace{-5mm}

\subsection{Detector families and their scalability}
\textbf{Formal verification.} Property checking, equivalence checking, and information‑flow verification (e.g., GLIFT/IFS) offer strong guarantees when specifications are precise and complete \cite{Tiwari2009GLIFT,Hu2011GLIFT,Nahiyan2017IFS}. The most visible downsides are (i) \emph{scalability}: state‑space growth and deep sequential logic make proofs expensive on industrial blocks; (ii) \emph{specification burden}: nontrivial effort is required to write and maintain comprehensive properties or security policies; (iii) \emph{reference dependence}: many flows assume a golden model, trusted micro‑architecture intent, or policy baselines that may be unavailable; and (iv) \emph{resolution}: passing a property check may not localize suspicious gates/nets, leaving engineers to manually triage large COIs once violations occur. These factors limit coverage for unseen Trojans that fall outside the encoded properties.

\textbf{Side-channel analyses.} Power, timing, electromagnetic, or thermal measurements can \emph{detect} HTs without full design visibility and are applicable \emph{pre-silicon} (simulation/emulation) and \emph{post-silicon} on fabricated chips; they do not imply inserting Trojans after fabrication \cite{Du2010CHES,Su2024NICE}. The most obvious drawbacks are (i) \emph{noise and variability}: process, voltage, temperature, and measurement noise reduce sensitivity and reproducibility; (ii) \emph{localization}: signatures are spatially coarse, especially in multi‑domain SoCs, making it hard to pinpoint malicious nodes; (iii) \emph{reference or calibration needs}: many techniques rely on golden chips, golden traces, or extensive calibration runs; and (iv) \emph{limited pre‑silicon utility}: modeling side‑channels accurately before tape‑out is challenging, which delays actionable feedback.

\textbf{Graph- and ML-based} detectors learn over netlists to classify nodes or logic COIs (e.g., GNN4TJ at RTL, gate-level GNNs, HW2VEC) \cite{Yasaei2021GNN4TJ,TrojanSAINT2023,Yu2021HW2VEC,Cheng2023GateDet}. They often need no golden reference, but face practical limits: large feature/edge tensors, multi-hop growth on high-fan-out/reconvergent logic, and distribution shift from small benchmarks to SoC-scale designs. Sampling can drop the long paths that link rare triggers to payloads; partitioning improves memory but may sever those paths or split COIs near rare nodes, undermining detection exactly where it matters.

\vspace{-3mm}

\subsection{SCOAP testability: metrics and rarity}
\label{subsec:scoap-rare-nodes}
Testability metrics quantify how difficult it is to control or observe internal signals. SCOAP (\emph{Sandia Controllability/Observability Analysis Program}) \cite{scoap} assigns three nonnegative integers to each net $n$: combinational controllability for logic 0, $\mathrm{CC0}(n)$ (effort to force logic~0), combinational controllability for logic 1, $\mathrm{CC1}(n)$ (effort to force logic~1), and combinational observability, $\mathrm{CO}(n)$ (effort to propagate $n$ to a primary output). By convention, primary inputs have $\mathrm{CC0}=\mathrm{CC1}=1$, and primary outputs have $\mathrm{CO}=0$. For a combinational network in topological order, common recurrences include:
\vspace{-5mm}

\[
\setlength{\jot}{1pt}
\begin{aligned}
\text{OR: }  & CC0(y)=\textstyle\sum_i CC0(x_i)+1,\\[-3pt]
             & CC1(y)=\min_i CC1(x_i)+1,\\[-3pt]
             & CO(x_i)=CO(y)+\!\!\sum_{j\neq i} CC0(x_j)+1,\\[-3pt]
\text{AND: } & CC1(y)=\textstyle\sum_i CC1(x_i)+1,\\[-3pt]
             & CC0(y)=\min_i CC0(x_i)+1,\\[-3pt]
             & CO(x_i)=CO(y)+\!\!\sum_{j\neq i} CC1(x_j)+1,\\[-3pt]
\text{INV: } & CC0(y)=CC1(x)+1,\;\; CC1(y)=CC0(x)+1,\\[-3pt]
             & CO(x)=CO(y)+1.
\end{aligned}
\]

For two‑input XOR/XNOR, controllabilities follow the usual pairwise minima of mixed sums; observabilities are analogous and depend on the easier of the two input values. Sequential circuits are handled by cutting at registers (treating flip‑flop outputs as pseudo‑inputs and inputs as pseudo‑outputs) and evaluating per cycle.

In practice, these integers are combined into a rarity score, for example:
\begin{equation}
\hspace{-5pt}R(n)=\max\{\text{CC0}(n),\text{CC1}(n)\}+\alpha\,\text{CO}(n),\;\; \alpha\!\in\![0.5,1],
\label{eq:rarity_formula}
\end{equation}
and nets are ranked by percentiles to isolate outliers. A \emph{higher} score indicates a net that is difficult to activate and/or difficult to propagate to an output (rarely activated under normal conditions), whereas a \emph{lower} score indicates a net that is comparatively easy to drive and observe (frequently activated). In this paper, the ranking guides the placement of observe/control points, the focus of specification-driven formal checks, and the emphasis of COIs during ML training.

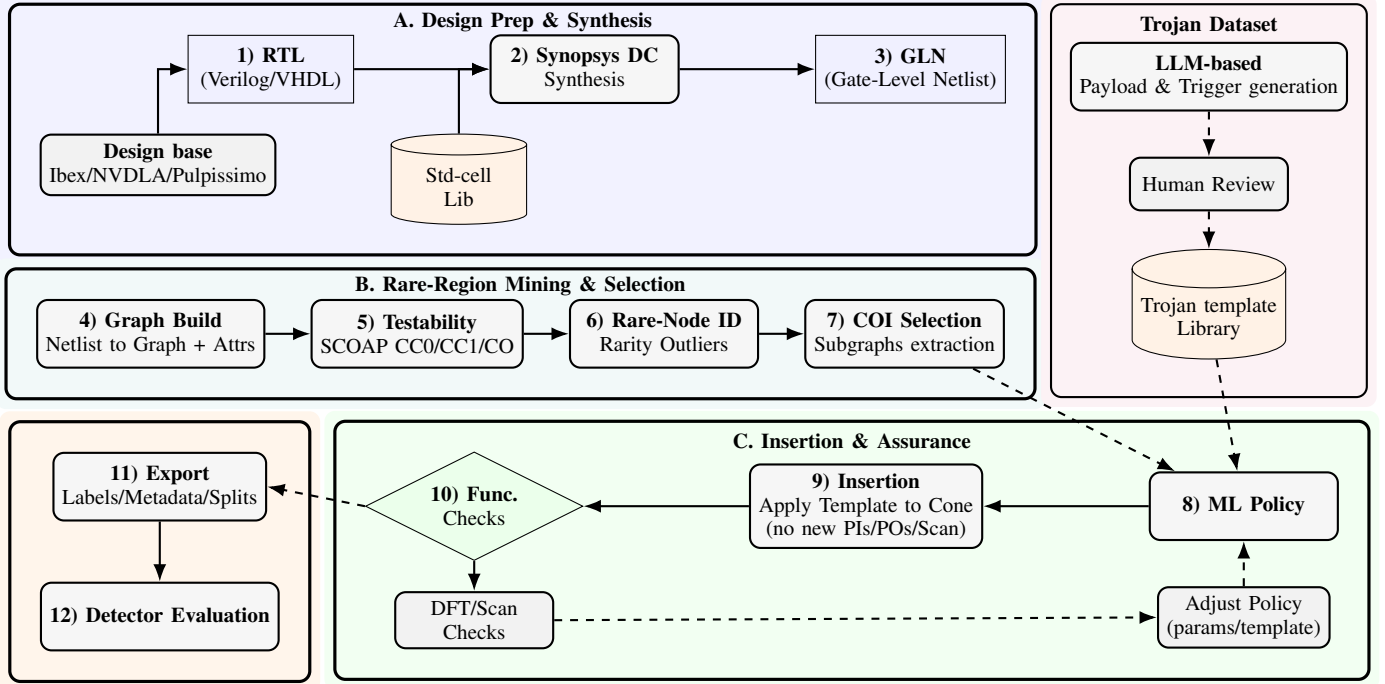
\begin{figure*}[t]
\centering

\begin{tikzpicture}[
    scale=0.2, 
    >=Latex,
    node distance=10mm and 10mm,
    font=\footnotesize,
    block/.style={draw, rounded corners, thick, align=center, fill=gray!7,
                  minimum width=25mm, minimum height=9mm},
    smallblock/.style={draw, rounded corners, thick, align=center, fill=gray!10,
                  minimum width=21mm, minimum height=7mm},
    doc/.style={draw, shape=rectangle, fill=blue!6,
                  minimum width=22mm, minimum height=9mm, align=center},
    db/.style={draw, shape=cylinder, shape border rotate=90, aspect=0.25,
                  fill=orange!10, minimum width=18mm, minimum height=12mm, align=center},
    decision/.style={draw, diamond, aspect=2, align=center, fill=green!8,
                       minimum height=9mm, minimum width=16mm},
    cloud/.style={draw, cloud, cloud puffs=12, cloud puff arc=90, aspect=2,
                  fill=purple!6, minimum width=26mm, minimum height=10mm, align=center},
    legend/.style={draw, rounded corners, thick, fill=white, align=left, inner sep=2mm},
    dashedarrow/.style={->, thick, dashed},
    arrow/.style={->, thick}
]

\node[doc] (rtl) {\textbf{1) RTL}\\(Verilog/VHDL)};
\node[smallblock, below=4mm of rtl, xshift=-15mm] (sdc) {\textbf{Design base} \\ Ibex/NVDLA/Pulpissimo};
\node[db,         below=4mm of rtl, xshift=+25mm] (lib) {Std-cell\\Lib};
\node[block, right=18mm of rtl] (dc) {\textbf{2) Synopsys DC}\\Synthesis};
\draw[arrow] (sdc) |- (rtl);
\draw[arrow] (rtl) -- (dc);
\draw[arrow] (lib) |- (dc);

\node[doc, right=18mm of dc] (gln) {\textbf{3) GLN}\\(Gate-Level Netlist)};
\draw[arrow] (dc) -- (gln);

\node[fit=(rtl)(sdc)(lib)(dc)(gln), draw, very thick, rounded corners, inner sep=4mm,
label={[yshift=-5mm]above:\textbf{A. Design Prep \& Synthesis}}] (grpA) {};

\node[block, below=26mm of gln, xshift=-101mm] (graph) {\textbf{4) Graph Build}\\Netlist to Graph + Attrs};

\node[block, right=6mm of graph] (scoap) {\textbf{5) Testability}\\SCOAP CC0/CC1/CO};
\node[block, right=6mm of scoap] (rare) {\textbf{6) Rare-Node ID}\\Rarity Outliers};
\node[block, right=6mm of rare]  (conesel) {\textbf{7) COI Selection}\\Subgraphs extraction};
\draw[arrow] (graph) -- (scoap);
\draw[arrow] (scoap) -- (rare);
\draw[arrow] (rare) -- (conesel);

\node[fit=(graph)(scoap)(rare)(conesel), draw, very thick, rounded corners, inner sep=4mm,
      label={[yshift=-5mm]above:\textbf{B. Rare-Region Mining \& Selection}}] (grpB) {};

\node[block,      right=8.5mm of gln, yshift=-1.mm] (llm)    {\textbf{LLM-based}\\Payload \& Trigger generation};
\node[smallblock, below=6mm  of llm]                 (Human) {Human Review};
\node[db,         below=5mm  of Human]              (tmpldb) {Trojan template\\Library};
\draw[dashedarrow] (llm) -- (Human);
\draw[dashedarrow] (Human) -- (tmpldb);

\node[fit=(llm)(Human)(tmpldb),
      draw, thick, rounded corners,
      inner xsep=2.5mm, inner ysep=5mm,
      label={[yshift=-5.5mm]above:\textbf{Trojan Dataset}}] (grpK) {};
\node[decision, below=11mm of graph, xshift=43mm] (cec) {\textbf{10) Func.}\\Checks};
\node[block, right=22mm of cec] (insert) {\textbf{9) Insertion}\\Apply Template to Cone\\(no new PIs/POs/Scan)};
\node[block, right=22mm of insert] (policy) {\textbf{8) ML Policy}};
\draw[arrow] (policy) -- (insert);
\draw[arrow] (insert) -- (cec);

\node[smallblock, below=4mm of cec] (dft) {DFT/Scan\\Checks};
\draw[arrow] (cec) -- (dft);

\draw[dashedarrow] (conesel) -- (policy);
\draw[dashedarrow] (tmpldb) -- (policy);

\node[smallblock, below=6mm of policy, align=center] (tune) {Adjust Policy\\(params/template)};
\draw[dashedarrow] (dft.east) -- (tune.west);
\draw[dashedarrow] (tune.north) -- (policy.south);
\node[fit=(policy)(insert)(cec)(dft)(tune), draw, very thick, rounded corners, inner sep=4mm,
      label={[yshift=-5mm]above:\textbf{C. Insertion \& Assurance}}] (grpC) {};

\node[block, left=13mm of cec, yshift=2.6mm] (export) {\textbf{11) Export}\\Labels/Metadata/Splits};
\node[block, below=8mm of export] (eval) {\textbf{12) Detector Evaluation}};
\draw[dashedarrow] (cec.west) -- (export.east);
\draw[arrow] (export) -- (eval);

\node[fit=(export)(eval), draw, very thick, rounded corners, inner sep=4mm,
      label={[yshift=4mm]above:\textbf{}}] (grpD) {};

\begin{scope}[on background layer]
  \node[fit=(grpA), fill=blue!5,   rounded corners] {};
  \node[fit=(grpB), fill=teal!5,   rounded corners] {};
  \node[fit=(grpK), fill=purple!5, rounded corners] {};
  \node[fit=(grpC), fill=green!5,  rounded corners] {};
  \node[fit=(grpD), fill=orange!8, rounded corners] {};
\end{scope}

\end{tikzpicture}

\caption{End-to-end pipeline. RTL and constraints are synthesized with \textbf{Synopsys DC} to a gate-level netlist (GLN). The GLN is converted to a graph with attributes; \textbf{SCOAP} yields testability scores used to identify \emph{rare} nodes and realistic COIs. A function-preserving \emph{template library} (populated via LLM-assisted trigger \& payload generation under human review) is combined with an \textbf{ML-based policy} to select placements and parameters. Inserted patterns must preserve functionality, \textbf{DFT/scan} checks, and \textbf{STA/area} budgets; failures trigger parameter readjustment. Successful instances are exported with labels/metadata/splits for reproducible detector evaluation. \emph{No new PIs/POs/scan behaviors are introduced at any stage.}}
\label{fig:pipeline}
\end{figure*}

\subsection{Scale gaps and computational bottlenecks}
Industrial designs are far larger than the academic benchmarks commonly used for evaluation. ISCAS’85/’89 and many Trust-Hub derivatives contain only $10^3$--$10^4$ gates, whereas production netlists routinely reach $10^6$--$10^8$ gates with millions of connections and severe class imbalance. Even static storage is substantial: a 32-dimensional feature matrix for 400\,k nodes is $\approx 49$\,MiB, and an edge list for 2.2\,M edges adds tens of MiB before accounting for model states or activations. 

\subsection{Need for unseen Trojans and diverse benchmarks}
Robust detection requires evaluation on \emph{unseen} trigger–payload combinations that differ statistically from training data and that reflect realistic class imbalance. A suitable benchmark dataset should preserve I/O functionality by construction for safe, comparable results; span a range of trigger sparsities and reconvergence depths; include benign payload equivalents that mimic footprint without harm; and provide precise per‑net and per‑cone labels for HT detectors. With such artifacts ML‑based detectors can be trained under realistic conditions rather than overfitting to a narrow family of Trojans.

\vspace{-5mm} 

\section{Methodology}
\label{sec:method}

This section details the end-to-end flow, as illustrated in Fig.~\ref{fig:pipeline}.
For each, the unmodified gate-level netlist (GLN) synthesized with \textit{Synopsys DC} serves as the golden baseline. Synthesized netlists and constraints are produced with standard combinational optimizations, preserving test structures. The resulting gate level netlists span from $\sim\!6.5\times 10^5$ to $\sim\!1.0\times 10^7$ edges across designs (see Table~\ref{tab:ingestion}), and the overall generation time---parsing through export---ranges from a median $\sim\!7$\,min at $\approx\!735$\,k edges to $\sim\!64.8$\,min at $\approx\!9.7$\,M edges, with memory rising near the $140$\,GB. Runtime and memory distribution across stages are shown in Table~\ref{tab:pipeline-runtime}.

\begin{table}[t]
  \caption{Scaling behavior of created designs.}
  \label{tab:ingestion}
  \centering
  \footnotesize
  \begin{tabular}{p{25pt}
    S[table-format=8.0, table-column-width=42pt]
    S[table-format=9.0, table-column-width=42pt]
    S[table-format=2.1, table-column-width=42pt]
    S[table-format=3.1, table-column-width=42pt]}
    \toprule
    {Design} & {\# of Nodes} & {\# of Edges} &
    {\shortstack{Parse\\Time [min]}} & {\shortstack{AST Peak\\RAM [GB]}} \\
    \midrule
    D1 & {30K}   & {184K}    & 1.8  & 2.3   \\
    D2 & {212K}  & {426K}    & 4.1  & 10.4  \\
    D3 & {120K}  & {735K}    & 7.1  & 29.1  \\
    D4 & {367K}  & {2.24M}   & 22.2 & 65.9  \\
    D5 & {883K}  & {4.95M}   & 42.1 & 120.5 \\
    D6 & {1.24M} & {7.42M}   & 53.2 & 131.7 \\
    D7 & {1.78M} & {9.68M}   & 64.7 & 140.5 \\
    \bottomrule
  \end{tabular}
\end{table}

\begin{figure}[b]
\centering
\resizebox{0.6\columnwidth}{!}{%
\begin{tikzpicture}[>=Latex]

  \tikzset{
    edge/.style={-, line width=0.28pt},
    nlow/.style={circle, draw, line width=0.3pt, minimum size=2.8mm, inner sep=0pt, fill=blue!22},
    nmed/.style={circle, draw, line width=0.3pt, minimum size=3.4mm, inner sep=0pt, fill=orange!55},
    nhigh/.style={circle, draw, line width=0.3pt, minimum size=4.4mm, inner sep=0pt, fill=red!50},
    nhighest/.style={circle, draw, line width=0.3pt, minimum size=4.4mm, inner sep=0pt, fill=red!100},
    lab/.style={font=\scriptsize}
  }
  \def\rInner{1.05}
  \def\rOuter{2.05}

  \node[nhighest] (C)  at (0,0)           {};

  \node[nhigh] (I1) at ( 90:\rInner)   {}; 
  \node[nhigh] (I2) at ( 30:\rInner)   {}; 
  \node[nmed]  (I3) at (-30:\rInner)   {}; 
  \node[nmed]  (I4) at (-90:\rInner)   {}; 
  \node[nmed]  (I5) at (-150:\rInner)  {}; 
  \node[nmed]  (I6) at ( 150:\rInner)  {}; 

  \node[nlow] (O1) at ( 90:\rOuter)    {}; 
  \node[nlow] (O2) at ( 30:\rOuter)    {}; 
  \node[nlow] (O3) at (-30:\rOuter)    {}; 
  \node[nlow] (O4) at (-90:\rOuter)    {}; 
  \node[nlow] (O5) at (-150:\rOuter)   {}; 
  \node[nlow] (O6) at ( 150:\rOuter)   {}; 

  \foreach \k in {1,...,6} { \draw[edge] (I\k) -- (C); }
  \draw[edge] (O1) -- (I1); \draw[edge] (O1) -- (I6);
  \draw[edge] (O2) -- (I2); \draw[edge] (O2) -- (I1);
  \draw[edge] (O3) -- (I3); \draw[edge] (O3) -- (I2);
  \draw[edge] (O4) -- (I4); \draw[edge] (O4) -- (I3);
  \draw[edge] (O5) -- (I5); \draw[edge] (O5) -- (I4);
  \draw[edge] (O6) -- (I6); \draw[edge] (O6) -- (I5);

  \node[lab] at ($(C)+( 0.55,0.00)$) {\texttt{N680}};
  \node[lab] at ($(I1)+( 0.3, 0.33)$) {\texttt{N35}};
  \node[lab] at ($(I2)+( 0.48, 0)$) {\texttt{N33}};
  \node[lab] at ($(I3)+( -0.48, -0.05)$) {\texttt{N683}};
  \node[lab] at ($(I4)+( -0.35,-0.25)$) {\texttt{N678}};
  \node[lab] at ($(I5)+(-0.48, 0.00)$) {\texttt{N685}};
  \node[lab] at ($(I6)+(-0.1, 0.35)$) {\texttt{N792}};

  \node[lab] at ($(O1)+( 0.48, 0)$) {\texttt{N301}};
  \node[lab] at ($(O2)+( 0.48, 0)$) {\texttt{N309}};
  \node[lab] at ($(O3)+( 0.48, 0)$) {\texttt{N305}};
  \node[lab] at ($(O4)+( 0.48,0)$) {\texttt{N294}};
  \node[lab] at ($(O5)+(-0.48, 0.00)$) {\texttt{N29}};
  \node[lab] at ($(O6)+(-0.42, 0.05)$) {\texttt{N401}};

  \node[lab, anchor=north] (Cinfo) at ($(C)+(-0.3,-2.20)$)
       {CC0/CC1/CO = 4/19/122 $\rightarrow$ $R{=}141$};
  \draw[dashed, line width=0.25pt] (C) -- (Cinfo.west);

  \node[lab, anchor=north] (O6info) at ($(O6)+(1.5,1.6)$)
       {CC0/CC1/CO = 1/2/10 $\rightarrow$ $R = 2 + 10 = 12$};
  \draw[dashed, line width=0.25pt] (O6) -- (O6info.west);

\end{tikzpicture}%
}
\caption{An example of rarity-annotated subgraph. Connections indicate edges that reconverge toward a central high-$R$ node. Nodes are labeled by net name; color encodes rarity $R$ (blue~$\rightarrow$~orange~$\rightarrow$~red for low~$\rightarrow$~medium~$\rightarrow$~high),   Example SCOAP values (calculated based on (\ref{eq:rarity_formula}) w/ $\alpha$ = 1) are shown for \texttt{N680}, and \texttt{N401}.}
\label{fig:rare-cone}
\end{figure}
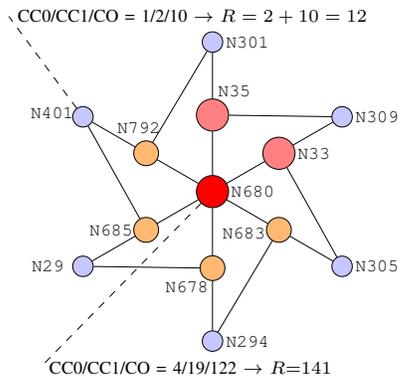

Following \emph{Graph Build} (Fig.~\ref{fig:pipeline}B-(4)), each GLN is converted into a directed connectivity graph whose nodes represent nets (optionally gates) and whose edges represent driver$\rightarrow$load relations. Lightweight structural attributes are attached to every node, including gate class, fan-in/fan-out, topological depth, local reconvergence signatures derived from input-cone overlap, and distances to design interfaces and sequential cut points. The flow then executes \emph{Testability/SCOAP} (Fig.~\ref{fig:pipeline}B-(5)) to compute $\mathrm{CC0}$, $\mathrm{CC1}$, and $\mathrm{CO}$ in linear time; rarity is ranked using the score $R(n)$ as defined in~(\ref{eq:rarity_formula}). 
\emph{Rare-Node ID} (Fig.~\ref{fig:pipeline}B-(6)) selects outlier nets by $R(\cdot)$ percentile and filters them by structural context typical of stealthy regions---such as reconvergent fan-in and branching into disjoint neighborhoods---so that candidates exhibit both low activity and realistic connectivity (see the stylized COI in Fig.~\ref{fig:rare-cone}). \emph{COI Selection} (Fig.~\ref{fig:pipeline}B-(7)) then extracts compact subgraphs around these rare nets, assembling/compiling per-subgraph metadata (size, depth to interfaces and registers, reconvergence descriptors) for the downstream policy.

Trigger–payload candidates are produced with an {\em Large-Language Model} (LLM). The model is used \emph{offline} to build a fixed dataset of stealthy trigger mechanisms and pass‑through payload behaviors that serve as reusable templates (the “Trojan Dataset” in Fig.~\ref{fig:pipeline}). The LLM prompt is designed to encourage diversity in both activation mechanisms and payload effect. On the trigger side, this includes patterns such as sequence detectors, counter windows, parity or Hamming checks, and inactivity- or timing-based anomalies. On the payload side, representative behaviors includes guarded multiplexers, shadow paths, inert togglers, and small offsets or bit flips. These representative families of trigger and payload mechanisms are summarized in Table~\ref{tab:trojan_samples_single_col}. The initial LLM-generated candidates are normalized into typed descriptors---trigger family, payload family, and parameter ranges such as tap count, local depth, permissible fan‑out growth, and reconvergence tolerance. Each descriptor is then \emph{compiled} into a local graph rewrite---a small, function‑preserving replacement in the gate level \textit{connectivity graph}. Concretely, within a bounded neighborhood an untriggered path is substituted with a trigger–payload template, while all primary inputs/outputs are left untouched. Each rewrite is validated with equivalence checking (following the flow~\cite{SynopsysFormalityUltraDS}): the “golden’’ and “trojanized’’ netlists are read, verification is restricted to the edited COI, compare points are mapped on the COI boundary, the trigger is constrained to its inactive value, and \texttt{verify} cmd is run to either prove equivalence or return a counterexample. Logic outside the COI is black-boxed to accelerate the check. 

\begin{table}[t]
\caption{Runtime and Peak RAM by Stage.}
\label{tab:pipeline-runtime}
\centering
\footnotesize
\begin{tabular}{l S[table-format=2.1] S[table-format=2.1] S[table-format=2.1]}
\toprule

{Stage} & {\shortstack{D3\\Time [min]}} &
{\shortstack{D4\\Time [min]}} & {\shortstack{Peak\\RAM [GB]}} \\

\midrule
Parsing/Graph Build         & 5.8 & 19.8 & 14.5 \\
SCOAP (forward/backward)    & 4.5 & 16.2 & 11.2 \\
Candidate Mining            & 2.2 &  7.0 &  6.8 \\
Template Insertion          & 1.9 &  6.2 &  4.1 \\
Functionality checks             & 1.3 &  2.8 &  3.6 \\
Labeling/Export             & 0.8 &  2.7 &  2.2 \\
{\bfseries Total (median)}           & 16.5 & 54.7 & 15.2 \\
\bottomrule
\end{tabular}
\end{table}

\begin{table}[b]
\centering

\vspace{-7mm}

\caption{Sample of Hardware Trojan Triggers and Payloads}
\label{tab:trojan_samples_single_col}
\resizebox{\columnwidth}{!}{%
\begin{tabular}{@{}ll@{}}
\toprule
\textbf{Name} & \textbf{Mechanism} \\
\midrule
\multicolumn{2}{c}{\textbf{Triggers}} \\
\midrule
3-State Sequence & Activates only when the FSM follows a specific, rare path. \\
Hamming Distance & Fires when input data is very close to a secret value. \\
Watchdog Timer & Activates from the prolonged absence of a specific event. \\
Glitch Detector & Triggers on a transient, single-cycle pulse on a signal line. \\
Hash-like Combo & Fires on a non-obvious combination of data and internal state. \\
\midrule
\multicolumn{2}{c}{\textbf{Payloads}} \\
\midrule
Timing Leak & Delays a critical output signal, leaking data via timing. \\
Off-by-One Error & Subtly corrupts data by adding one to an output value. \\
FSM Deadlock & Forces the state machine into a valid state but prevents it from leaving. \\
Counter Drift & Causes a sampling counter to miscount, slowly degrading operation. \\
Reset Disable & Prevents a key register from resetting, corrupting future operations. \\
\bottomrule
\end{tabular}%
}
\end{table}

The \emph{ML Policy} is implemented as a compact multilayer perceptron (MLP) with a two-input design that decides \emph{where} to place a given template of triggers and payloads. For every candidate, the pipeline constructs (i) a feature vector that summarizes the \emph{subgraph of rare nets}, including its rarity percentiles, reconvergence indicators, and distances between design elements, and (ii) a feature vector that encodes the \emph{template choice and parameters}, including trigger/payload pair, number of taps, and the intended local insertion pattern. These two vectors are concatenated and passed through the MLP, which outputs two interpretable scores: an \emph{acceptance score} estimating whether the proposed candidate will pass the required checks, and a \emph{stealthy score} estimating how statistically challenging the instance will be for detectors. A simple weighted combination ranks all candidates. The MLP is trained in a self-supervised manner using the pipeline’s own history: candidates that passed all checks become positive examples; candidates that failed equivalence or DFT/scan checks become structured negatives. Over time, the policy learns to prefer subgraphs and parameterizations that are both likely to be accepted and measurably hard for detectors, without requiring a golden reference model.

\begin{figure}[t]
    \centering
    \includegraphics[width=\columnwidth, height=6cm, keepaspectratio]{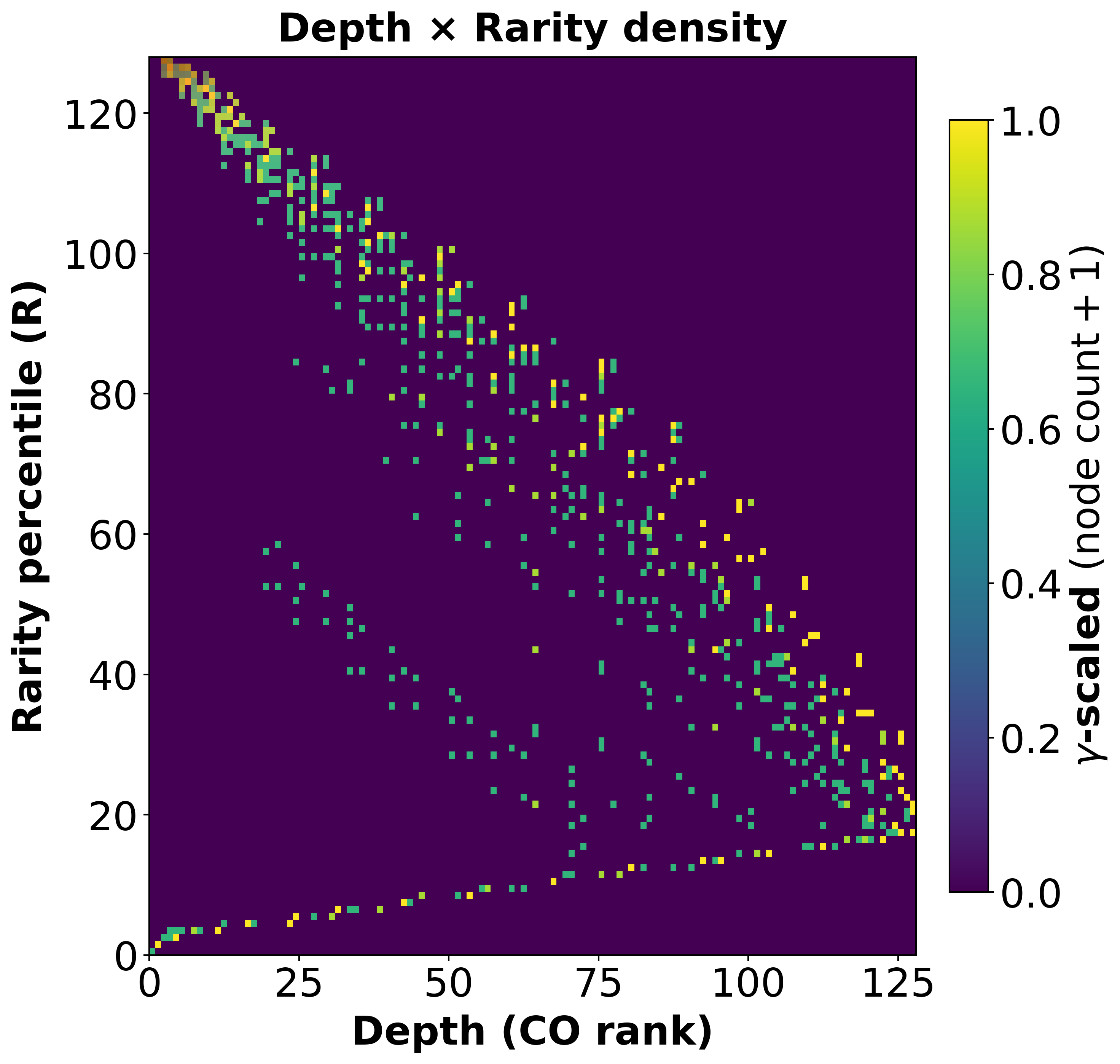}
    \caption{Nodes rarity heatmap for a SoC with many shallow logic COIs (the \textit{Ibex RISC-V CPU} core). As expected, set of rare nodes shifts to the \emph{right} because deeper nets (higher CO) usually raise $R$. Yet, extremely rare nets can be identified at modest depth (upper-left) when controllability is very poor.}
    \vspace{-5pt}
    \label{fig:depth_rarity}
\end{figure}

\emph{Insertion} (Fig.~\ref{fig:pipeline}C-(9)) applies the selected template as a local graph rewrite (deterministic, template-driven splice) inside the chosen subgraph of rare nets. The tool extracts a small cone of influence (bounded by flip-flops) around a rare net and inserts a parameterized trigger–payload template whose \emph{inactive branch is a pass-through} (i.e., it leaves the signal unchanged, equivalent to a wire). Every insertion is subjected to checks to ensure \emph{functionally unperturbed} modification (forcing trigger to be inactive; setting \textit{scan$\_$en, test$\_$mode} to 0) (Fig.~\ref{fig:pipeline}C-(10)); any mismatch triggers automatic rollback and negative feedback to the policy. In parallel, \emph{DFT/Scan Checks} verify that scan chains, test enables, and observability/controllability of scan cells are unchanged; candidates that alter test behavior are rejected and recorded as policy negatives. 

Ideally, Trojan triggers should be inserted deep inside the logic path and must be activated rarely. However, there are examples of SoCs, where the logic is broken into many short, simple stages separated by registers (\textit{Ibex RISC-V CPU} core). Figure~\ref{fig:depth_rarity} shows that scenario: nodes rarity grows with depth because combinational observability, CO, contributes directly to $R(n)=\max\{\text{CC0}(n),\text{CC1}(n)\}+\alpha\,\text{CO}(n),\; (\alpha = 1) $, yet the upper-left cluster reveals nets that are extremely rare even at modest depth when controllability is very poor. This reveals an insertion trade-off. While targeting the rarest trigger might lead to a logic COI of minimal depth that is controlled by a long, weakly observable path, pursuing greater logic depth increases activation rarity via CO at the risk of incurring significant timing and routing overhead. 

Finally, the \emph{Export} stage assembles/compiles each successful instance into a reproducible benchmark: the “Trojanized” GLN, including per-net attributes and metadata, and fixed train/validation/test splits at the subgraph (COI) level. Since the insertions are function-preserving by construction and approved by equivalence and DFT/scan checks, the resulting set provides industrial-scale, challenge instances that reflect realistic rarity and reconvergence, enabling empirical, head-to-head evaluation of detection methods under the same class imbalance and scale observed in modern SoCs.

\section{Experimental Results}
\label{sec:eval}

The evaluation focuses on whether state-of-the-art detectors trained on modest, publicly available designs can reliably identify Trojans embedded in much larger, structurally different designs with \emph{unseen} trigger--payload characteristics. 
Insertions follow the methodology in Section~\ref{sec:method}: rare regions are identified by SCOAP-based rarity, compact subgraphs are extracted, and function-preserving modifications are applied using fixed  trigger and payload templates (Table~\ref{tab:trojan_samples_single_col}). 

Open-source implementations of representative graph and ML-based detectors are selected to reflect the strongest publicly documented baselines. The study includes the following Trojan detectors.
\begin{itemize}[leftmargin=10pt,labelindent=1pt,labelsep=0.5em,itemsep=2pt,topsep=2pt]
  \item \textbf{HW2VEC} (HOST'21)~\cite{Yu2021HW2VEC}: an automated framework that extracts RTL or gate-level graphs (AST/DFG/GLN) and feeds a graph-classification backend. In experimental setup, the backend uses a \emph{GNN4TJ} graph classifier~\cite{Yasaei2021GNN4TJ} following the reference repository configuration file: \cite{HW2VEC_GNN4TJ_Config}. 
  \item \textbf{Netwise} (TCAD’24)~\cite{utyamishev2024netwise}: a gate-level, node-wise detector that embeds local neighborhoods into “embedding clouds” and applies scalable convolutions to reduce memory pressure and improve generalization at SoC scale.
  \item \textbf{NetVGE} (TCAD’25)~\cite{popryho2025netvge}: a KGE+transformer pipeline over weighted variable-dependency graphs that combines an entity transformer (local interactions) with a context transformer (global dependencies) and weighted self-attention emphasizes rarely accessed nodes.
\end{itemize}
Models are trained on the TrustHub dataset \cite{trusthub} that contains both clean netlists and their Trojanized counterparts. 

The evaluation is performed on the proposed larger-scale designs, with samples shown in Table \ref{tab:ingestion}. These designs are held out from the training and validation sets to ensure a realistic test of generalization, as it mirrors the real-world scenario where a model must identify unseen Trojans in a target SoC whose architecture and its characteristics are not present during training.
The ability of the baseline framework to process the generated benchmarks and detect Trojans is shown in Fig.~\ref{fig:capability_matrix}. In terms of netlist parsing, all detectors successfully handle designs D1--D3. However, only the KGE-based methods, \emph{NetVGE} \cite{popryho2025netvge} and \emph{Netwise} \cite{utyamishev2024netwise}, scale to D4--D7, whereas the GNN-based solution fails to process the largest netlists. 

\subsection{HW2VEC performance at SoCs scale}
HW2VEC with the GNN4TJ model is designed to operate on data-flow graphs and aggregate data over local neighborhoods. When applied to gate-level SoCs, achieving a receptive field large enough to connect deep, low-activity triggers to their corresponding payloads requires traversing numerous hops through high fan-out, reconvergent logic. In such case, the resulting activation/memory footprint grows superlinearly in edge count, and practical mitigation techniques (e.g., neighbor sampling, hard partitioning) sever precisely the long-range dependencies needed for rare-predicate reasoning. On D1, HW2VEC shows capability to detect presence of rarely triggered nodes in design, but does not correctly identify the trigger COI. The sensitivity of the HW2VEC detector to rare nodes drops sharply at D2 and becomes statistically indistinguishable from random chance in the 99--100\% rarity tail. The results shows that these constraints manifest as (i) parse/memory failures on D4--D7, and (ii) a collapse of rare-tail recall starting at D2. The distributional gap between TrustHub-scale training graphs and multi-million-edge test graphs further amplifies this degradation.

\subsection{NetVGE/Netwise performance on large gate‑level graphs}

In contrast to the GNN model, both KGE-based methods retain \emph{rare‑tail} sensitivity by design. \emph{Netwise} aggregates local neighborhoods into embedding clouds and applies scalable set of convolutions, reducing reliance on very deep message passing and preserving rare‑node signal through mid‑scale designs. As a result, its sensitivity degrades on D4–D7, where triggers and payloads are separated by long, rarely activated, and reconvergent paths within the control and datapath logic.

\begin{figure}[!t]
    \centering
\includegraphics[width=\columnwidth]{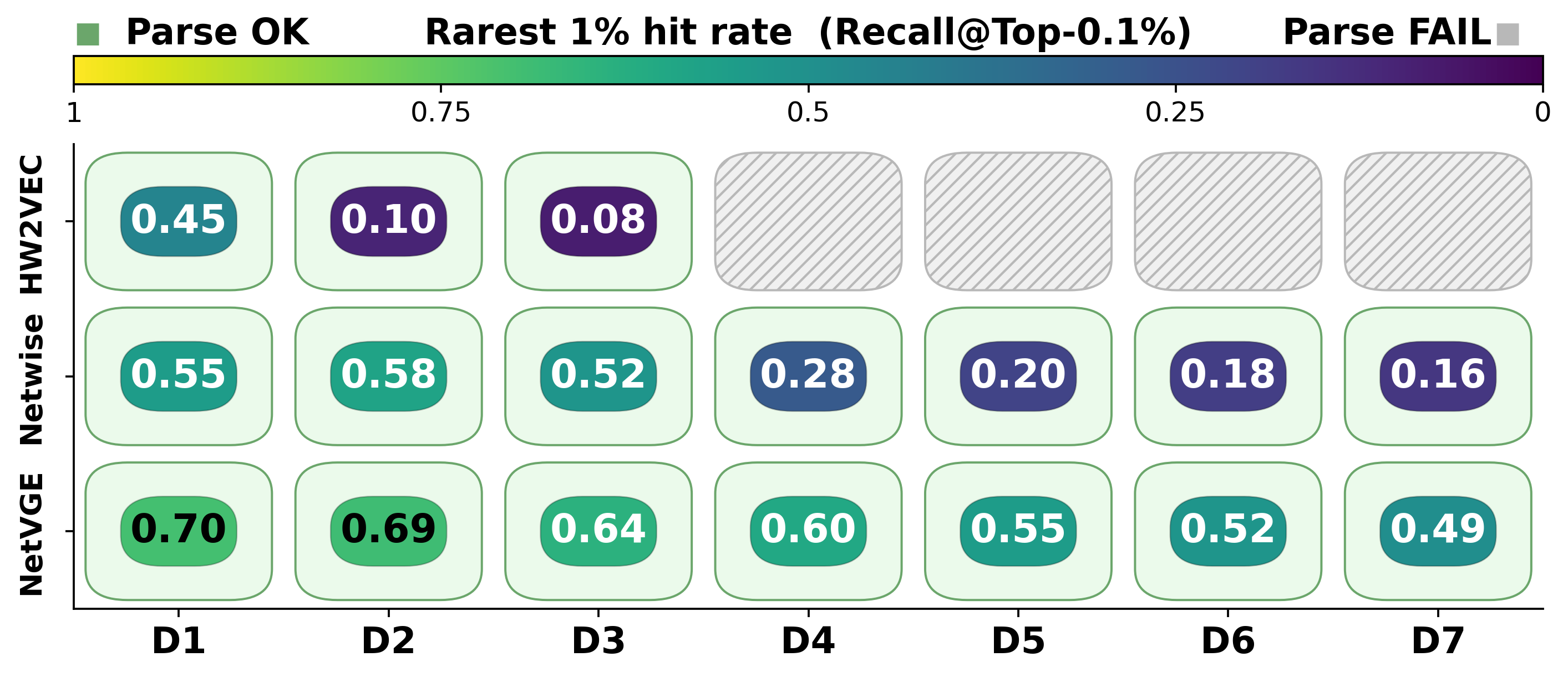}
    \vspace{-4mm}
    \caption{HT detectors capability matrix}
    \vspace{-5mm}
    \label{fig:capability_matrix}
\end{figure}

The \emph{NetVGE} method encodes weighted variable-dependency tuples, embeds them with KGE, and scores them in latent space using a hierarchical transformer. This architecture consists of an entity transformer for local interactions and a context transformer for global dependencies, with a weighted self-attention mechanism that emphasizes rarely accessed nodes. The approach was evaluated on designs D2–D7, where it demonstrated the ability to detect triggers located in the rarest circuit locations (99\textsuperscript{th} percentile). As shown in Fig.~\ref{fig:capability_matrix}, it continues to localize affected logic regions at a level where other detectors fail to scale, although overall accuracy remains modest (49\%-70\%).

Both detectors succeed in highlighting \emph{portions} of the affected COIs (rare nodes near the trigger gates) but fall short of localizing the precise trigger logic, underscoring the difficulty of long-range dependencies tracking under extreme class imbalance.

\vspace{5mm}
\subsection{Key observations}

Based on the experiments, general-purpose GNNs cannot reliably detect \emph{unseen}, stealthy Trojans at SoC scale when trained on much smaller publicly available benchmarks. This limitation is not only a matter of data volume. Linking rare triggers to their downstream effects requires a long-range view of the circuit, which conflicts with practical memory limits and the graph partitioning strategies used to manage them. These strategies, while necessary, disrupt the logical pathways that a detector must follow. The proposed artifact set therefore serves a dual purpose: it provides SoC-scale challenge instances with \emph{unseen} Trojans, and it highlights the need for model designs that reason hierarchically over large graphs, incorporate testability-aware signals, and preserve long-range relationships without incurring prohibitive resource costs.

\vspace{-0.99mm}
\section{Conclusion}
\label{sec:conclusion}

An automated, scalable methodology is presented for inserting Trojan-like, function‑preserving patterns into industrial‑scale netlists to stress‑test HT detectors under conditions that mirror modern SoCs. The flow converts large gate‑level netlists into graphs, mines statistically rare regions using SCOAP testability, and applies local rewrites drawn from a fixed LLM-generated library of trigger–payload templates that leave user‑visible behavior unchanged and pass equivalence and DFT/scan checks. A lightweight placement policy, trained on pipeline feedback, selects where to instantiate templates so that accepted insertions are both feasible and statistically stealthy. The resulting benchmarks are labeled in a self-supervised at the net and subgraph levels, enabling reproducible evaluations at scale.

Empirical results indicate that representative state‑of‑the‑art graph‑based detectors, trained on smaller public designs, do not generalize to unseen, industrial‑size targets populated with unseen trigger–payload families. Precision–recall performance collapses as rarity and reconvergence increase, and attempts to preserve long‑range context through a deeper aggregation clash with practical memory and runtime limits. These observations suggest the gap is architectural rather than merely data‑driven: detectors must learn to reason over large graphs hierarchically, incorporate testability‑aware structure, and retain long‑range dependencies without prohibitive resource growth. 

The proposed methodology provide a controlled and scalable  way to synthesize realistically stealthy challenge instances while keeping I/O behavior intact, narrowing the evaluation gap between academic benchmarks and production‑scale designs. Limitations remain: rarity mining relies primarily on SCOAP with sequential cuts, payloads are pass‑through analogs rather than truly malicious effects, and experiments focus on a set of open designs. Future work will extend rarity analysis with sequential and path‑sensitization signals, broaden template diversity within the same function‑preserving envelope, and investigate hierarchical, testability‑aware models and hybrid formal/learning strategies that can scale to the connectivity and imbalance regimes of modern SoCs.

\clearpage
\IEEEtriggeratref{12}
\bibliographystyle{IEEEtran}
\bibliography{references}

@inproceedings{scoap,
  author    = {Lawrence H. Goldstein and Evelyn L. Thigpen},
  title     = {{{SCOAP}: Sandia Controllability/Observability Analysis Program}},
  booktitle = {Proceedings of the 17th Design Automation Conference (DAC)},
  year      = {1980},
  pages     = {190--196},
  publisher = {ACM},
  doi       = {10.1145/800139.804528}
}

@article{adatest,
  author    = {Huili Chen and Xinqiao Zhang and Ke Huang and Farinaz Koushanfar},
  title     = {{{AdaTest}: Reinforcement Learning and Adaptive Sampling for On-chip Hardware {Trojan} Detection}},
  journal   = {ACM Transactions on Embedded Computing Systems},
  year      = {2023},
  volume    = {22},
  number    = {2},
  articleno = {37},
  numpages  = {23},
  doi       = {10.1145/3544015}
}

@inproceedings{rltestgen,
  author    = {Zhixin Pan and Prabhat Mishra},
  title     = {{Automated Test Generation for Hardware {Trojan} Detection using Reinforcement Learning}},
  booktitle = {Proceedings of the 26th Asia and South Pacific Design Automation Conference (ASPDAC '21)},
  year      = {2021},
  pages     = {408--413},
  publisher = {ACM},
  doi       = {10.1145/3394885.3431595}
}

@inproceedings{gnn4gate,
  author    = {Dong Cheng and Chen Dong and Wenwu He and Zhenyi Chen and Yi Xu},
  title     ={ {{GNN4Gate}: A Bi-Directional Graph Neural Network for Gate-Level Hardware {Trojan} Detection}},
  booktitle = {Design, Automation \& Test in Europe Conference (DATE)},
  year      = {2022},
  pages     = {1315--1320},
  note      = {EDAA}
}

@article{nhtdgl,
  author  = {Kento Hasegawa and Kazuki Yamashita and Seira Hidano and Kazuhide Fukushima and Kazuo Hashimoto and Nozomu Togawa},
  title   = {{Node-Wise Hardware {Trojan} Detection Based on Graph Learning}},
  journal = {IEEE Transactions on Computers},
  year    = {2025},
  volume  = {74},
  number  = {3},
  pages   = {749--761},
  doi     = {10.1109/TC.2023.3280134}
}

@misc{trusthub,
  author    = {H. Salmani and M. Tehranipoor and R. Karri},
  title     = {{{Trust-Hub} Hardware {Trojan} Benchmarks}},
  howpublished = {\url{https://www.trust-hub.org}},
}

@inproceedings{iscas85,
  author    = {Franc Brglez and Hideo Fujiwara},
  title     = {{A Neutral Netlist of 10 Combinational Benchmark Circuits and a Targeted Translator in {FORTRAN}}},
  booktitle = {Proceedings of the IEEE International Symposium on Circuits and Systems (ISCAS)},
  year      = {1985},
  address   = {Kyoto, Japan}
}

@inproceedings{Tiwari2009GLIFT,
  author    = {Mohit Tiwari and Xun Li and Hassan M. G. Wassel and Frederic T. Chong and Timothy Sherwood},
  title     = {{Complete Information Flow Tracking from the Gates Up}},
  booktitle = {Proceedings of the 14th International Conference on Architectural Support for Programming Languages and Operating Systems (ASPLOS)},
  pages     = {109--120},
  year      = {2009},
  publisher = {Association for Computing Machinery}
}

@article{Hu2011GLIFT,
  author  = {Wei Hu and Jason Oberg and Mohit Tiwari and Timothy Sherwood and Ryan Kastner},
  title   = {{Gate-Level Information Flow Tracking for Security Lattices}},
  journal = {ACM Transactions on Design Automation of Electronic Systems},
  volume  = {20},
  number  = {1},
  pages   = {2:1--2:25},
  year    = {2014},
  doi     = {10.1145/2629350}
}

@inproceedings{Nahiyan2017IFS,
  author    = {Adib Nahiyan and Mehdi Sadi and Rahul Vittal and Gustavo Contreras and Domenic Forte and Mark Tehranipoor},
  title     = {{Hardware {Trojan} Detection through Information Flow Security Verification}},
  booktitle = {Proceedings of the IEEE International Test Conference (ITC)},
  year      = {2017},
  publisher = {IEEE}
}

@inproceedings{Du2010CHES,
  author    = {Min Du and Fei Li and Ye Li and Kang Wei},
  title     = {{A Scalable Hardware {Trojan} Detection Methodology for {CUTs} Using Statistical Analysis of Side-Channel Information}},
  booktitle = {Cryptographic Hardware and Embedded Systems -- CHES 2010},
  series    = {Lecture Notes in Computer Science},
  volume    = {6225},
  pages     = {328--342},
  publisher = {Springer},
  year      = {2010},
  doi       = {10.1007/978-3-642-15031-9_12}
}

@inproceedings{Su2024NICE,
  author    = {Ting Su and Yaohua Wang and Shi Xu and Lusi Zhang and Simin Feng and Jialong Song and Yiming Liu and Yongkang Tang and Yang Zhang and Shaoqing Li and Yang Guo and Hengzhu Liu},
  title     = {{Improving the Ability of Thermal Radiation Based Hardware {Trojan} Detection via Noise-Induced Pixel Occupation Enhancement ({NICE})}},
  booktitle = {Proceedings of the 33rd {USENIX Security} Symposium ({USENIX Security} 2024)},
  address   = {Philadelphia, PA, USA},
  month     = {Aug},
  year      = {2024},
  publisher = {USENIX Association},
  url       = {https://www.usenix.org/conference/usenixsecurity24/presentation/su-ting}
}

@inproceedings{Yasaei2021GNN4TJ,
  author    = {Rozhin Yasaei and Shih{-}Yuan Yu and Mohammad Abdullah Al Faruque},
  title     = {{{GNN4TJ}: Graph Neural Networks for Hardware {Trojan} Detection at Register Transfer Level}},
  booktitle = {Proceedings of the 2021 Design, Automation \& Test in Europe Conference \& Exhibition (DATE)},
  year      = {2021}
}

@INPROCEEDINGS{TrojanSAINT2023,
  author={Lashen, Hazem and Alrahis, Lilas and Knechtel, Johann and Sinanoglu, Ozgur},
  booktitle={2023 IEEE International Symposium on Circuits and Systems (ISCAS)}, 
  title={{{TrojanSAINT}: Gate-Level Netlist Sampling-Based Inductive Learning for Hardware {Trojan} Detection}}, 
  year={2023},
  volume={},
  number={},
  pages={1-5},
  doi={10.1109/ISCAS46773.2023.10181403}}

@inproceedings{Yu2021HW2VEC,
  author    = {Shih{-}Yuan Yu and Rozhin Yasaei and Qingrong Zhou and Tommy Nguyen and Mohammad Abdullah Al Faruque},
  title     = {{{HW2VEC}: {A} Graph Learning Tool for Automating Hardware Security}},
  booktitle = {2021 IEEE International Symposium on Hardware-Oriented Security and Trust (HOST)},
  pages     = {13--23},
  year      = {2021},
  publisher = {IEEE},
  doi       = {10.1109/HOST49136.2021.9702281}
}

@article{Cheng2023GateDet,
  author  = {Dong Cheng and Chao Dong and Wei He and Zhiqiang Chen and Xiaoyang Liu and Hui Zhang},
  title   = {{A Fine-Grained Detection Method for Gate-Level Hardware {Trojan} Based on Bidirectional Graph Neural Networks ({GateDet})}},
  journal = {Journal of King Saud University -- Computer and Information Sciences},
  year    = {2023},
  note    = {In press}
}

@article{utyamishev2024netwise,
  author   = {D. Utyamishev and I. Partin-Vaisband},
  title    = {{{Netwise} Detection of Hardware {Trojans} using Scalable Convolution of Graph Embedding Clouds}},
  journal  = {IEEE Transactions on Computer-Aided Design of Integrated Circuits and Systems},
  volume   = {43},
  number   = {10},
  pages    = {3116--3128},
  year     = {2024},
  month    = {oct},
}

@article{popryho2025netvge,
  title   = {{{NetVGE}: {Netwise} Hardware {Trojan} Detection at {RTL} Using Variable Dependency and Knowledge Graph Embedding}},
  author  = {Popryho, Yaroslav and Pal, Debjit and Partin-Vaisband, Inna},
  journal = {IEEE Transactions on Computer-Aided Design of Integrated Circuits and Systems},
  year    = {2025},
  publisher = {IEEE}
}

@online{SynopsysPrimeTimeHyperScale2025,
  title   = {{{STA} Strategies for Fast and Efficient Signoff in Multi-Billion Instance Designs}},
  author  = {{Synopsys}},
  year    = {2025},
  url     = {https://www.synopsys.com/articles/sta-strategies-multi-billion-instance-designs.html},
  note    = {Describes PrimeTime HyperScale for hundreds of millions to billions of instances}
}

@online{SynopsysFormalityUltraDS,
  author   = {{Synopsys, Inc.}},
  title    = {{{Formality} and {Formality Ultra} Datasheet}},
  year     = {2024},
  url      = {https://www.synopsys.com/content/dam/synopsys/verification/datasheets/formality-and-formality-ultra-ds.pdf},
  urldate  = {2025-09-14},
}

@online{HW2VEC_GNN4TJ_Config,
  author   = {{AICPS Lab}},
  title    = {{{HW2VEC} \texttt{example\_gnn4tj.yaml}  configuration}},
  year     = {2021},
  url      = {https://github.com/AICPS/hw2vec/blob/master/examples/example_gnn4tj.yaml},
  urldate  = {2025-09-14},
}

\end{document}